\documentclass[
    ,final            % use final for the camera ready runs
    ,numberedheadings % uncomment this option for numbered sections
  ]
  {aipproc}

\layoutstyle{6x9}

\begin{document}

\newcommand{\ez}{\mbox{\.z}}
\newcommand{\en}{\mbox{\'n}}
\newcommand{\el}{\mbox{\l{}}}

\title{Using COSY--11 apparatus for the precise studies of the natural width of the $\eta'$ meson}



\author{Eryk Czerwi{\en}ski}{
  address={Institute of Physics, Jagiellonian University, Poland
  \& IKP Research Centre J\"ulich, Germany}
}

\author{Pawe{\el} Moskal for the COSY--11 collaboration}{
  address={Institute of Physics, Jagiellonian University, Poland
  \& IKP Research Centre J\"ulich, Germany}
}

\begin{abstract}
We present preliminary results and motivation of measurement of the total
width of the $\eta'$ meson ($\Gamma_{\eta'}$).
\end{abstract}

\maketitle

\section{Importance of $\Gamma_{\eta'}$}
The physics of the $\eta'$ meson receives an increasing interest in view of the forthcoming
measurements planned e.g. at the COSY, DA$\phi$NE-2 and MAMI-C facilities where the $\eta'$ will be
produced in hadron-hadron, e$^{+}$-e$^{-}$, and $\gamma$-hadron reactions, respectively.

Experimentally the emphasis will be put on the studies of the $\eta^{\prime}$ meson decays
which are of interest on its own account and certainly will provide inputs to the phenomenology
of the Quantum Chromo-Dynamics~\cite{bugra} in the non-perturbative regime.
Specifically, precise determinations of the partial widths for the $\eta^{\prime}$ decay
channels will be helpful for the development of the Chiral Perturbation Theory.
However, the experimental precision of the partial width for various decay channels
 -- where only the branching ratio is known or will be measured --
is governed by  the precision of the knowledge of the total width.
In the case of the $\eta'$ meson the branching ratios are typically known with
precision better than 1.5\%, while the total width is established
about 10 times less accurate~\cite{pdg}.
Therefore, we expect that the precise determination of the natural width of the $\eta'$ meson  will have
a strong impact on the physics results which will be derived from 
measurements carried out by collaborations:
WASA-at-COSY~\cite{hhadam}, CBall-at-MAMI~\cite{thomas} and KLOE-2~\cite{ambrosino,caterina}.

\section{Previous experiments}
In the last issue of the Review of Particle Physics
only two direct measurements of the natural width of the $\eta'$ meson
are reported~\cite{pdg}. In the first experiment the width
was established from the missing mass spectrum
of the $\pi^- p \to n X $ reaction measured close to the threshold for
the production of the $\eta^{\prime}$ meson~\cite{binnie}.
The experimental mass resolution achieved was equal to 0.75~MeV/c$^2$~(FWHM)
and the extracted value of $\Gamma_{\eta^{\prime}}$ amounts to $0.28~\pm~0.10$~MeV/c$^2$.
In the second experiment the value of $\Gamma_{\eta^{\prime}}$ was derived
from the threshold excitation function of the $pd\to ^3\!\!He~X$  reaction~\cite{wurzinger}.
The study was performed at 20 different beam momenta, however,
the error of the $\Gamma_{\eta^{\prime}}$ was larger than in the previous measurement
due to the large relative monitoring uncertainities. In this experiment
$\Gamma_{\eta^{\prime}}~=~0.40\pm0.22$~MeV/c$^2$ was determined.
The mean value of the width of the $\eta'$ meson from the two
direct measurements~\cite{binnie,wurzinger}
amounts to \mbox{(0.30$\pm$0.09)~MeV/c$^{2}$}~\cite{pdg}
and differs from the value of \mbox{(0.202$\pm$0.16)~MeV/c$^{2}$}
determined indirectly from the combinations of partial widths obtained from
integrated cross sections and branching ratios~\cite{pdg}. The significantly
more precise direct determination of the $\Gamma_{\eta'}$ should resolve
this discrepancy.

\section{New measurement}
\label{we}
\begin{figure}[h]
  \includegraphics[width=0.37\textwidth]{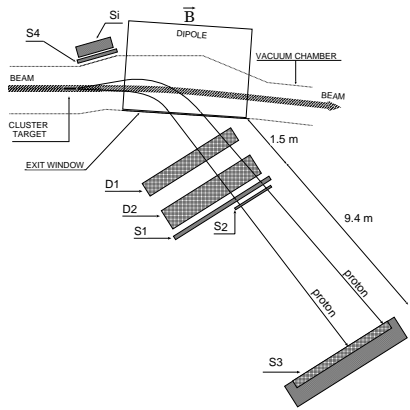}
  \caption{COSY--11 detection setup. D1 and D2 denote drift chambers used
           for reconstruction of trajectories of positively charged ejectiles. 
           (S1-S2-S3) scintillator hodoscope is used for the time of flight determination. 
           Elasticaly scattered protons
           are registered by silicon pad (Si) and scintillator (S4) detectors (recoil protons)
           and by drift chambers and S1 counter (forward protons).}
  \label{jeden}
\end{figure}

Based on many years of experience gained with the COSY-11 apparatus~\cite{brauksiepe,pawel}
we expect that 
combined with the excellent features of the stochastically cooled proton
beam of COSY~\cite{prasuhn,stockhorst} it will allow to significantly reduce the present uncertainty 
in the value of the natural width of the $\eta'$ meson.
The $\Gamma_{\eta'}$ will be derived directly from
the missing mass distribution of the $pp \to ppX$ reaction measured near
the kinamatical threshold.  The advantage of a study close to the threshold
is that the uncertainities of the missing mass determination
are much reduced since $\partial(mm)\slash\partial p$ tends to zero. Here by \emph{mm} we denoted
the missing mass and by \emph{p} the momentum of the outgoing protons.
The experiment was conducted for five
discret values of beam momentum: 3211, 3213, 3214,
3218, and 3224~MeV\slash c, where
the threshold value amounts to 3208.3~MeV\slash c.
In order to improve the experimental resolution
of the four-momentum determination of the registered particles
and in order to decrease the spread
of the momentum of the beam protons reacting with the target
two major changes have been applied to the COSY-11 setup~(Fig.~\ref{jeden}).
Namely,
the spatial resolution of the measurement of the particle track coordinates in
the drift chambers was improved by
increasing  the supply voltage up to  the maximum allowed value and also
the dimensions of the target in the direction perpendicular to the COSY beam was decreased
from 9 to circa 1~mm.
The changes were possible due to the development of the M\"unster type cluster-jet
target~\cite{dombrowski,taeschner}.

Due to the high value of the dispersion at  the position of the COSY-11 target,
the decrease of the size of the reaction volume resulted in a significant reduction 
of the momentum spread seen by the target. The effect is visualized in Fig.~\ref{piec}a.
As can be clearly implied from the figure, the information about target size 
is crucial for determination of beam momentum spread. It is also of great importance
for the estimation of an error of momentum reconstruction
of outgoing protons.  Therefore, in order to control the systematical uncertainties
each of the crucial parameters (like target dimesions
or beam momentum spread) were monitored by at least two independent methods.

\subsection{Target - a crucial element}
\begin{figure}[h]
  \makebox[0pt]{
   \raisebox{0.38\textheight}{
    \makebox[-0.91\textwidth][c]{}
    \makebox[0.0\textwidth][c]{{\bf a)}}
   }
  }
  \makebox[0pt]{
   \raisebox{0.18\textheight}{
    \makebox[-0.91\textwidth][c]{}
    \makebox[0.0\textwidth][c]{{\bf d)}}
   }
  }
  \makebox[0pt]{
   \raisebox{0.38\textheight}{
    \makebox[-0.22\textwidth][c]{}
    \makebox[0.0\textwidth][c]{{\bf b)}}
   }
  }
  \makebox[0pt]{
   \raisebox{0.18\textheight}{
    \makebox[-0.24\textwidth][c]{}
    \makebox[0.0\textwidth][c]{{\bf e)}}
   }
  }
  \makebox[0pt]{
   \raisebox{0.00\textheight}{
    \makebox[0.70\textwidth][c]{}
    \makebox[0.00\textwidth][c]{\includegraphics[width=0.28\textwidth]{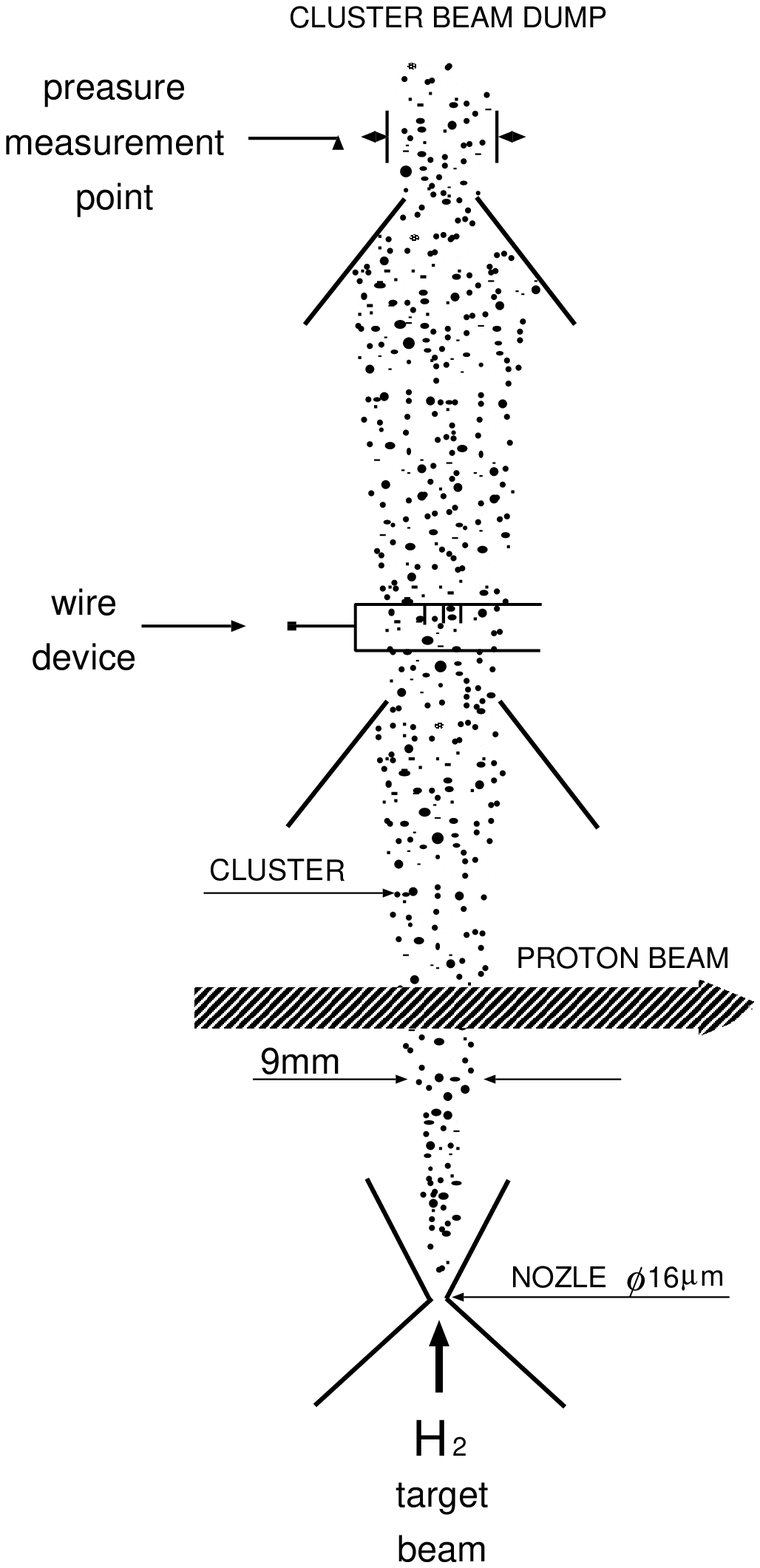}}
   }
  }
  \makebox[0pt]{
   \raisebox{0.20\textheight}{
    \makebox[0.088\textwidth][c]{}
    \makebox[0.00\textwidth][c]{\includegraphics[width=0.27\textwidth]{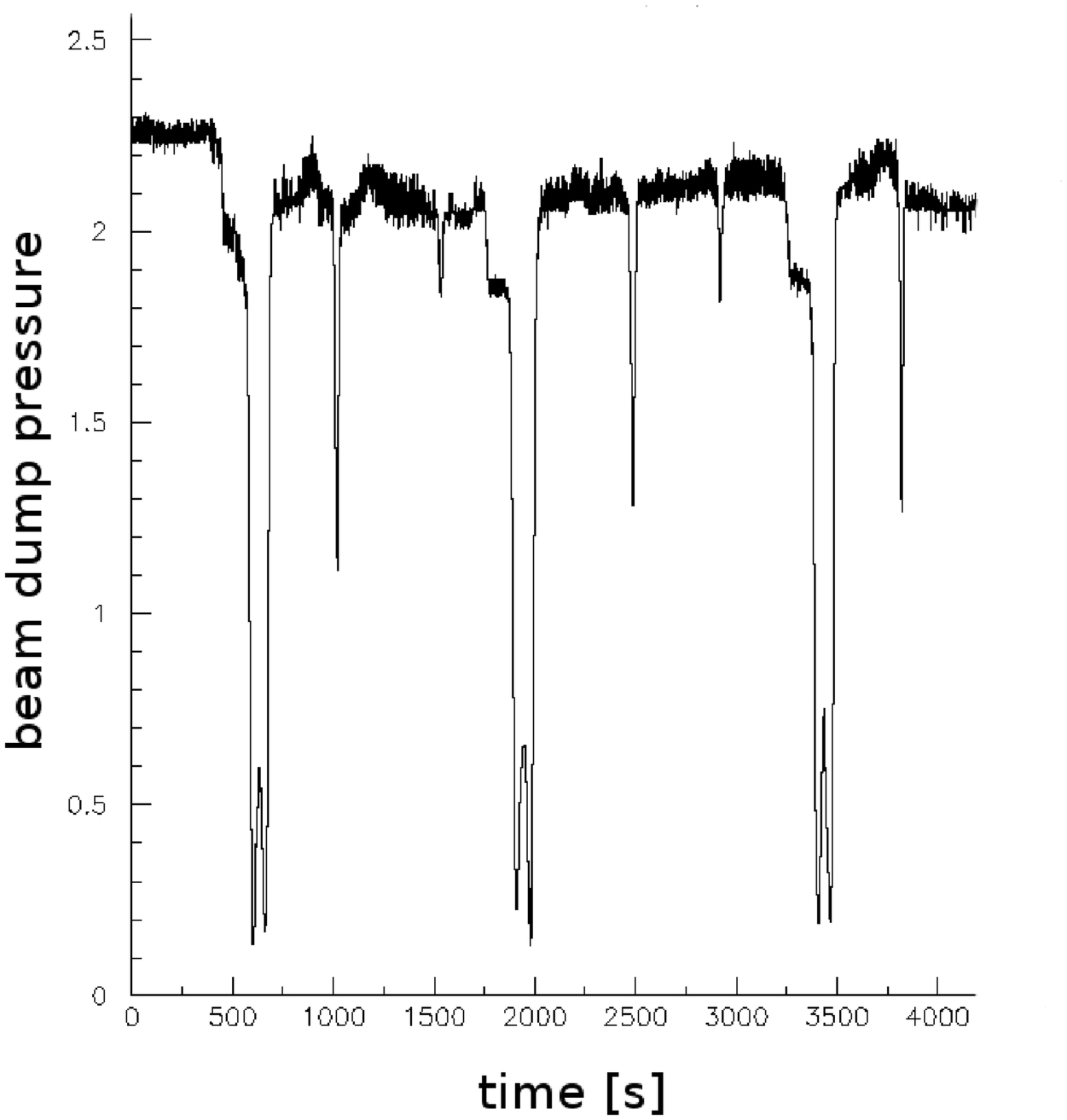}}
   }
  }
  \makebox[0pt]{
   \raisebox{0.00\textheight}{
    \makebox[-0.64\textwidth][c]{}
    \makebox[0.00\textwidth][c]{\includegraphics[width=0.31\textwidth]{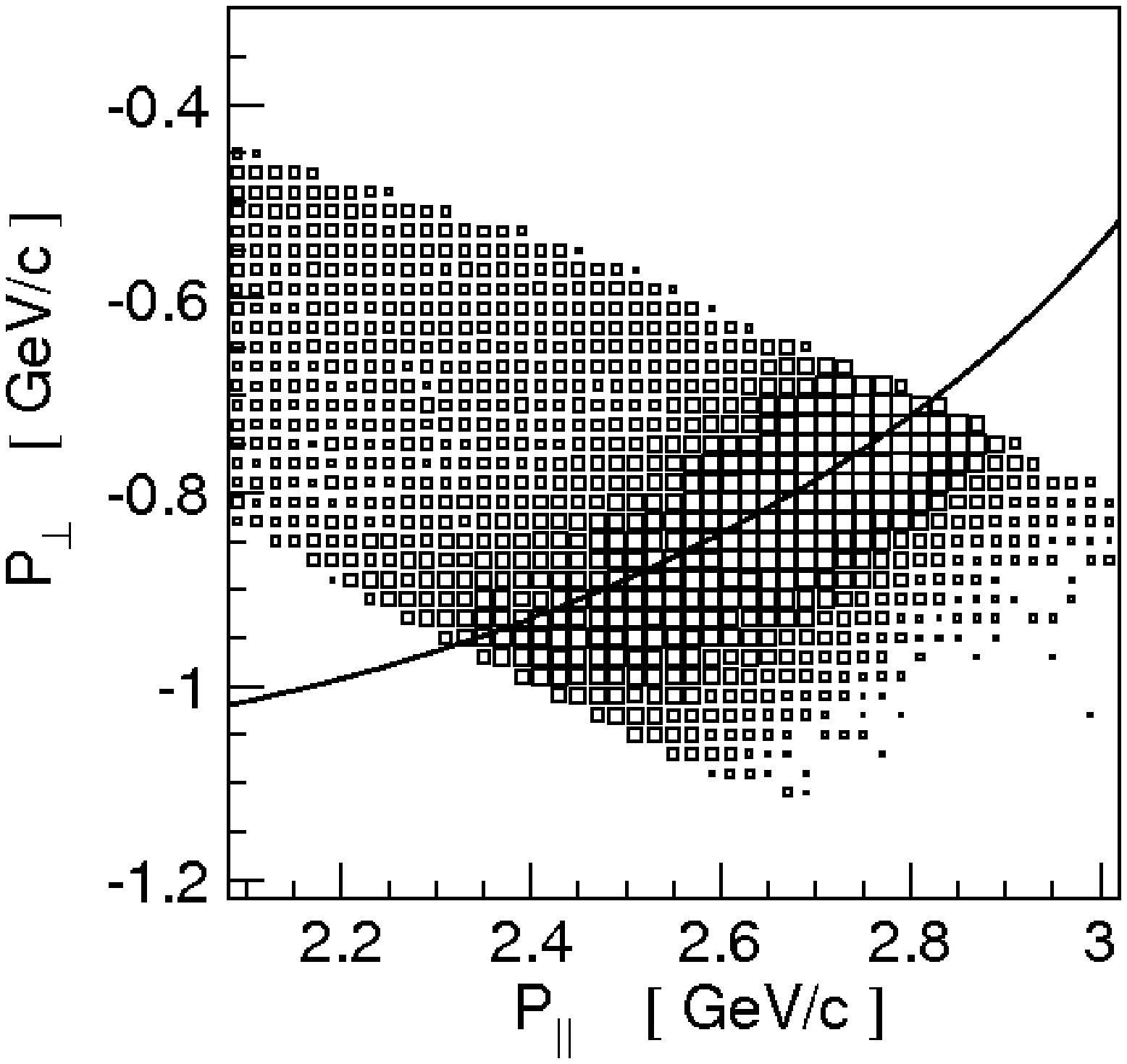}}
   }
  }
  \makebox[0pt]{
   \raisebox{0.005\textheight}{
    \makebox[0.02\textwidth][c]{}
    \makebox[0.00\textwidth][c]{\includegraphics[width=0.27\textwidth]{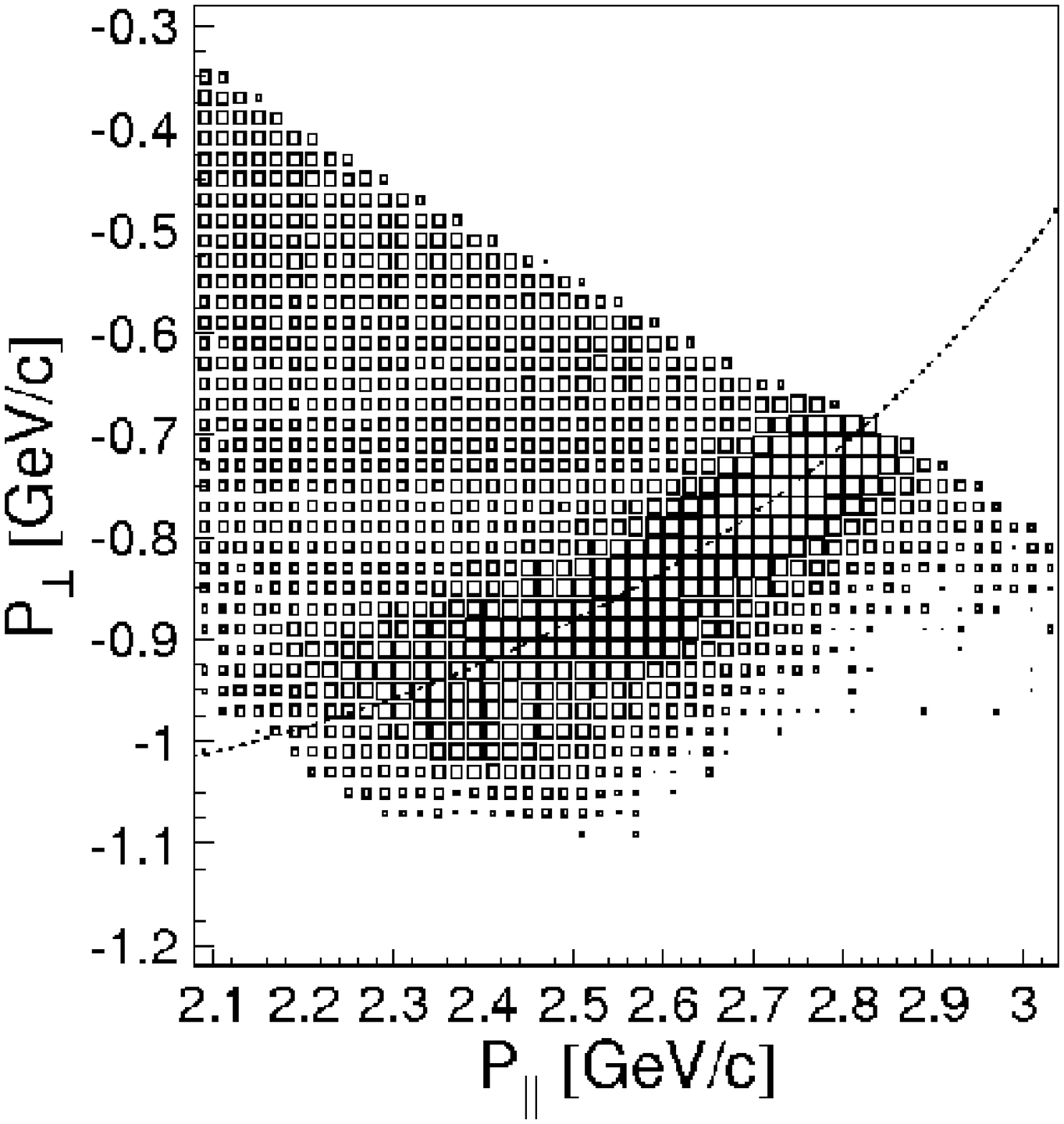}}
   }
  }
  \makebox[0pt]{
   \raisebox{0.20\textheight}{
    \makebox[-0.67\textwidth][c]{}
    \makebox[0.00\textwidth][c]{\includegraphics[width=0.30\textwidth]{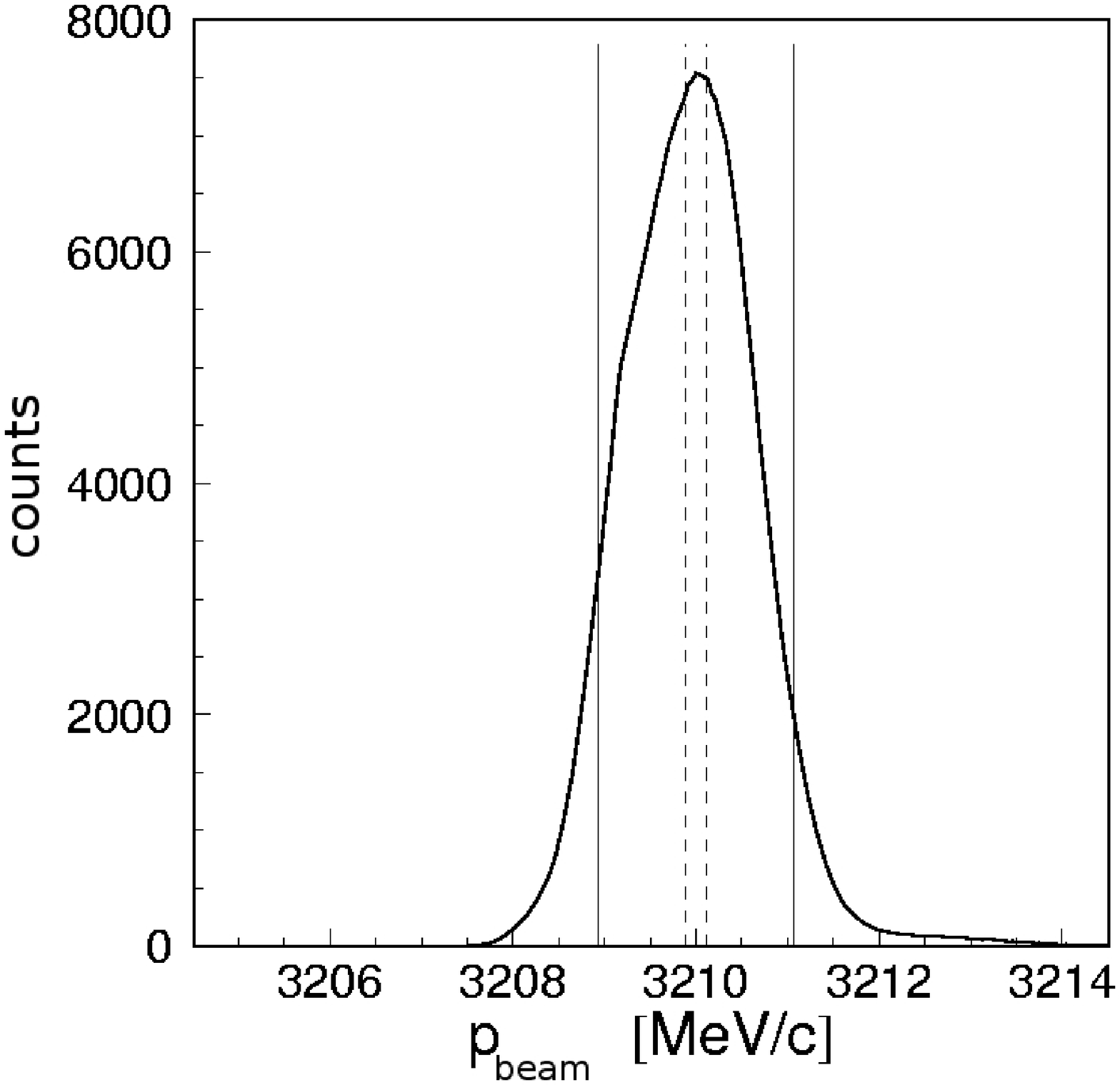}}
   }
  }
  \makebox[0pt]{
   \raisebox{0.20\textheight}{
    \makebox[0.35\textwidth][c]{}
    \makebox[0.0\textwidth][c]{{\bf c)}}
   }
  }
  \caption{
           a) Beam momentum distribution obtained from the Schotky frequency spectrum
           measured during one of the previous
           COSY-11 runs. The range "seen" by 9~mm and 1~mm
           target is marked by the solid and  dashed lines respectively.
           b) Distribution of the pressure measured during the wire device rotation.
           c) Schematic view of the of the target and beam crossing.
           d) and e): Distribution of the elastically scattered protons.
           Number of entries per bin is shown in logarithmic scale.
           Superimposed lines depict expected kinematical ellipses.
           d) One of the previous COSY-11 experiments
           (target width: 9~mm, $p_{beam}=$2010~MeV/c~\cite{rafal})
           e) On-line data from the reported here experiment
           (target width: circa 1~mm, $p_{beam}=$3211~MeV/c)}
  \label{piec}
\end{figure}

The systematical
error of the extraction of $\Gamma_{\eta'}$ will depend on the accuracy of the determination
of the missing mass
resolution. This depends predominantly on the knowledge of the momentum spread of the COSY beam
(Fig.~\ref{piec}a) and on the
accuracy of the four-momentum determination of the registered protons.
In the case of the experimental technique used by the COSY-11 collaboration,
both mentioned factors, depend on the dimensions of the target.
Therefore it is crucial to monitor precisely the spatial  size
of the target perpendicular to the COSY beam.
To this end we exploit two independent methods.
One of the techniques relies on the measurement of the gas pressure
in the last chamber of the cluster-jet dump (Fig.~\ref{piec}b-c),
and the second method is based on the determination
of the momentum distribution of elasticaly scattered protons (Fig.~\ref{piec}d-e).
The gas pressure was measured as a function of the position of wires
moving with the constant velocity through the cluster beam.
These wires were rotated around the axis perpendicular to the beam of hydrogen clusters.
In case that one or more wires crossed the target the clusters hitting
a wire were stopped causing a decrease of the pressure in the last stage
of the cluster dump.
The wire device was located above the reaction point (Fig.~\ref{piec}c)
allowing to monitor the target dimensions
without disturbing the  measurement of the $pp\to pp\eta'$ reaction.
The knowledge of the wire thickness and geometry enables
to calulate the target dimensions from the ditribution of the pressure 
as a function of the position of the device.
Fig.~\ref{piec}b shows an example of the measured 
pressure as function of time using a constant angular velocity of the device.
The extraction of the target dimensions is in progress. From a preliminary analysis 
we expect to achieve an accuracy of about 0.2~mm.

The second technique used for monitoring a target spatial size is based on 
the measurement of the momentum distribution of the elastically scattered protons.
The momentum reconstruction of registered protons is performed by tracing back trajectories
from  drift chambers through the dipole magnetic field to the
target, which is assumed to be an infinitely thin vertical line.
In reality, however, the reactions take place in that region of finite
dimensions where beam and target overlap.
Consequently, assuming in the analysis an infinitesimal target implies a
smearing of the momentum vectors and hence a decrease of the resolution of the
missing mass signal.
Therefore, we have developed a method to monitor this overlap by measuring the elastically
scattered protons~\cite{pawel}.
Trajectories of the protons scattered in the forward direction are
measured by means of two drift chambers (D1 and D2) and a scintillator hodoscope~(S1),
whereas the recoil protons are registered in coincidence with the
forward ones using a silicon pad detector arrangement (Si) and a scintillation detector~(S4).
The two--body kinematics gives an unambiguous relation between the
scattering angles of the recoiled and forward flying protons.  Therefore, the events of
elastically scattered protons can be identified from the correlation
line formed between the position in the silicon pad detector Si, and
the scintillator hodoscope S1,
the latter measured by the two drift chamber stacks.  For those
protons which are elastically scattered in forward direction and  are deflected
in the magnetic field of the dipole the momentum vector at the target point
can be determined.
According to two--body kinematics, the momentum components parallel and
perpendicular to the beam axis form an ellipse.  An example is shown in Fig.~\ref{piec}d-e.
The width of the distributions can be used as a measure of the size of the interaction region.
For the appraisal of the effect in Fig.~\ref{piec}d-e
we present results obtained with targets of 9 and 1 millimeters.

\subsection{Systematic error}
Measurement of the missing mass distributions at five different beam energies
will allow for monitoring the systematic uncertainities in the determination 
of the experimental mass resolution.
This is mainly because the smearing of the missing mass due to the natural width of the $\eta^{\prime}$
remains unaltered when the beam momentum changes, whereas
the smearing caused by the experimental uncertainities
will narrow with the decreasing  beam momentum and
at threshold it will reach a constant value directly proportional to the
spread of the beam momentum.  The effect is shown in Fig.~\ref{szesc}a. 
This figure illustrates also that the change of the target thickness
by 8~mm results in a change of the mass resolution by about 0.3~MeV. Since we expect to control
the target thickness with the accuracy better then 0.2~mm, the systematical error due to the determination
of the target size should be smaller than 0.01~MeV.

Moreover, we can also control the influence on the mass resolution caused by a different experimental sources.
For example from the angular distribution of the missing mass spectrum 
we will be able to estimate contributions to the mass resolution due to the spread of the beam momentum
and due to the proton momentum reconstruction. This is because 
the resolution of the missing mass due to the spread of the beam momentum
is almost independent of the polar emission angle
of the $\eta^{\prime}$ meson, whereas the smearing of  the missing mass due to the
uncertainty of the proton momentum reconstruction does depend on
the emission angle of the $\eta^{\prime}$ meson. 

\subsection{Preliminary results}
\begin{figure}[t]
    \makebox[0pt]{
     \makebox[0.33\textwidth][c]{
      \makebox[0.11\textwidth][c]{}
      \makebox[0.33\textwidth][c]{\includegraphics[width=0.32\textwidth]{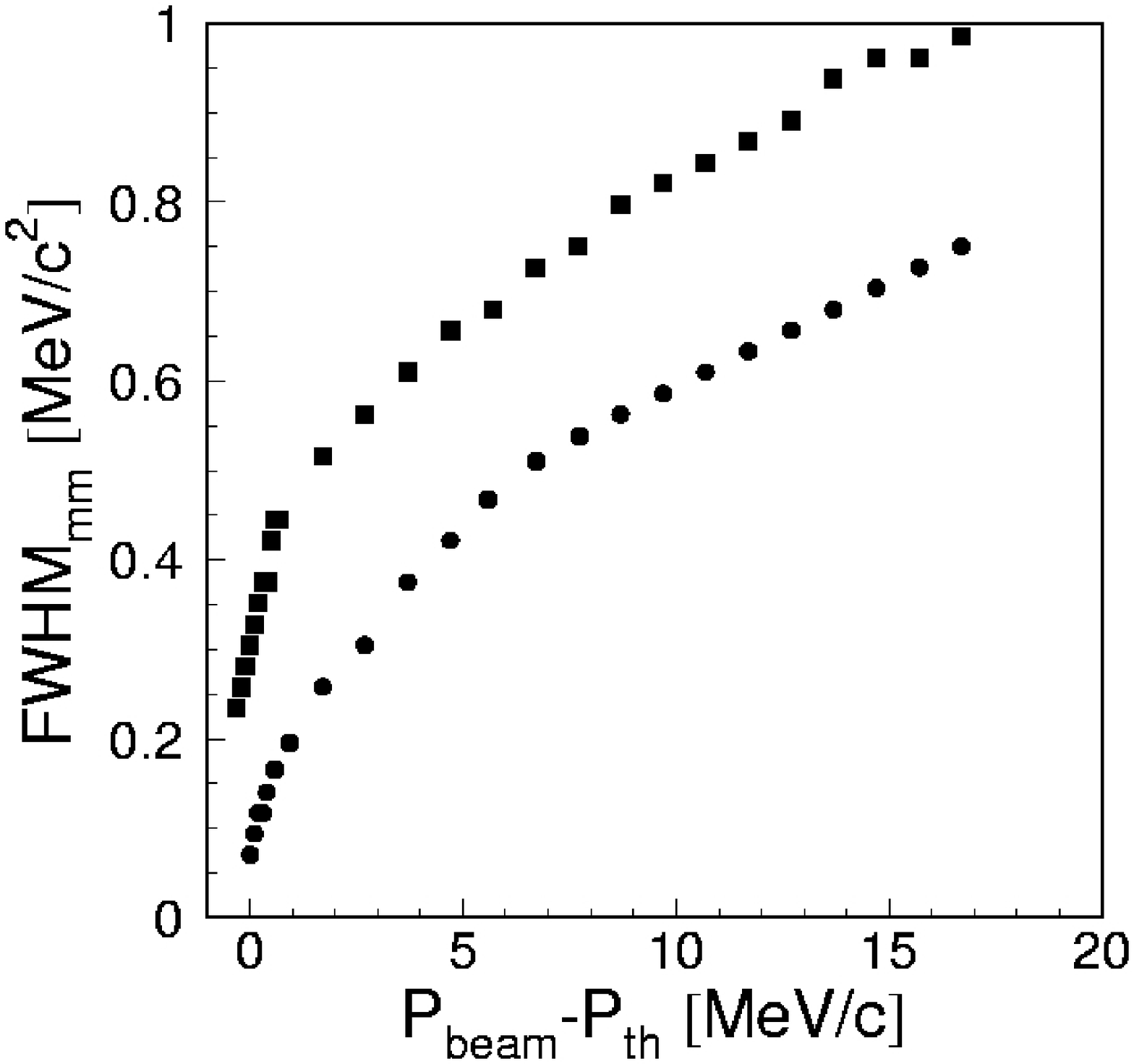}}
     }
     \raisebox{0.002\textheight}{
      \makebox[0.33\textwidth][c]{
       \makebox[0.06\textwidth][c]{}
       \makebox[0.33\textwidth][c]{\includegraphics[width=0.25\textwidth]{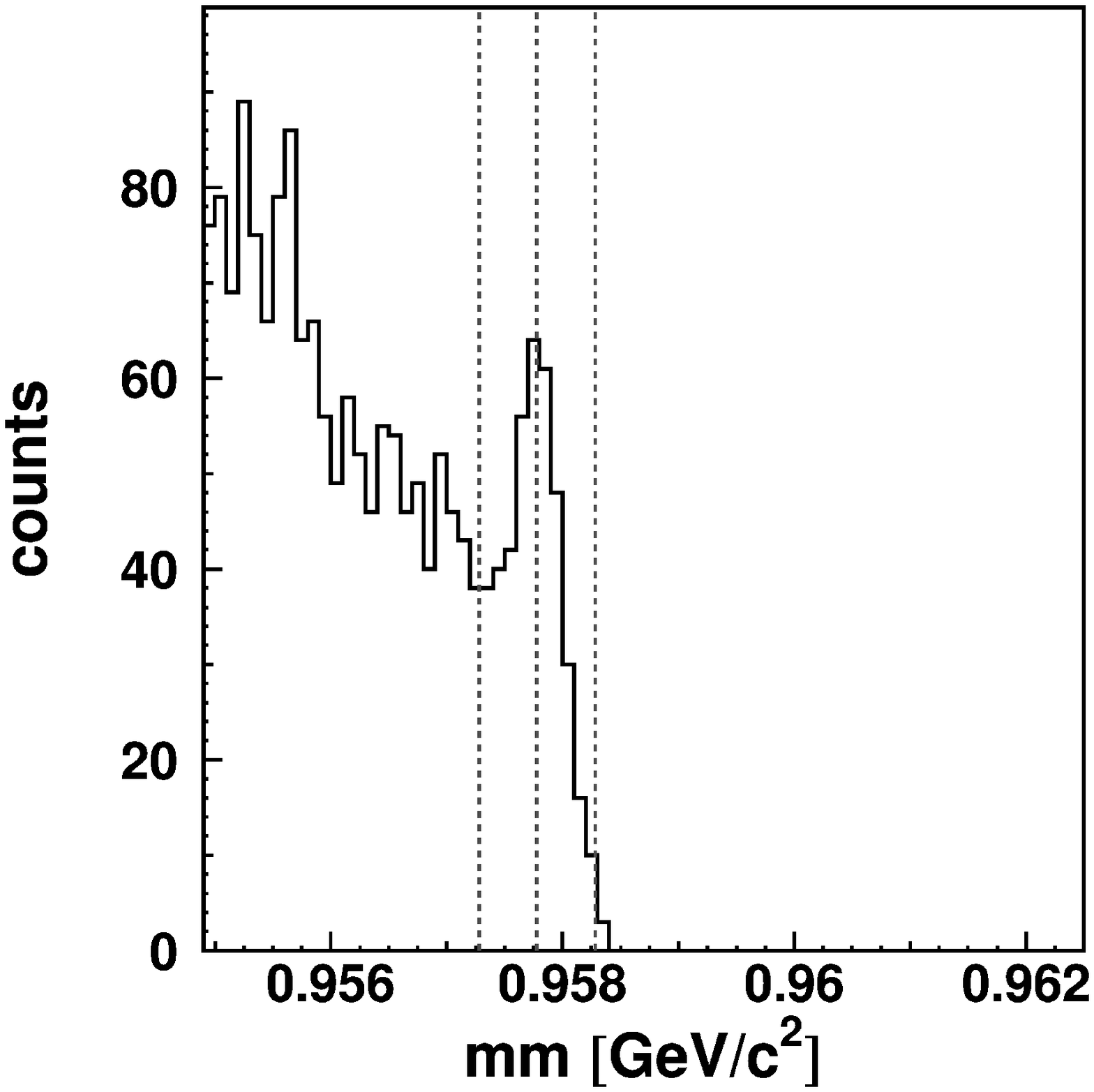}}
      }
      \makebox[0.33\textwidth][c]{
       \makebox[0.05\textwidth][c]{}
       \makebox[0.33\textwidth][c]{\includegraphics[width=0.25\textwidth]{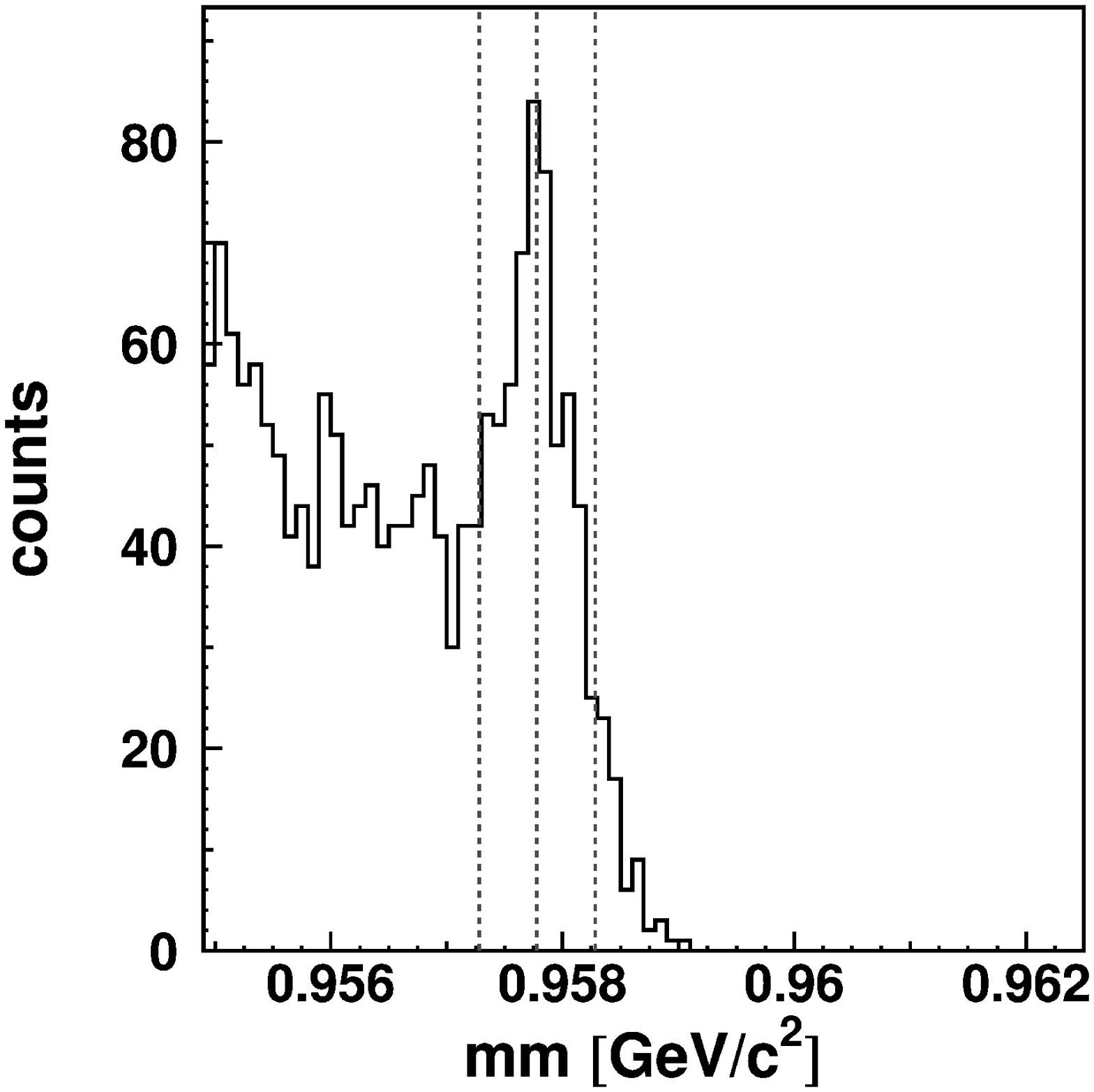}}
      }
     }
    }
    \makebox[0pt]{
     \raisebox{-0.17\textheight}{
      \makebox[0.33\textwidth][c]{
       \makebox[0.06\textwidth][c]{}
       \makebox[0.33\textwidth][c]{\includegraphics[width=0.25\textwidth]{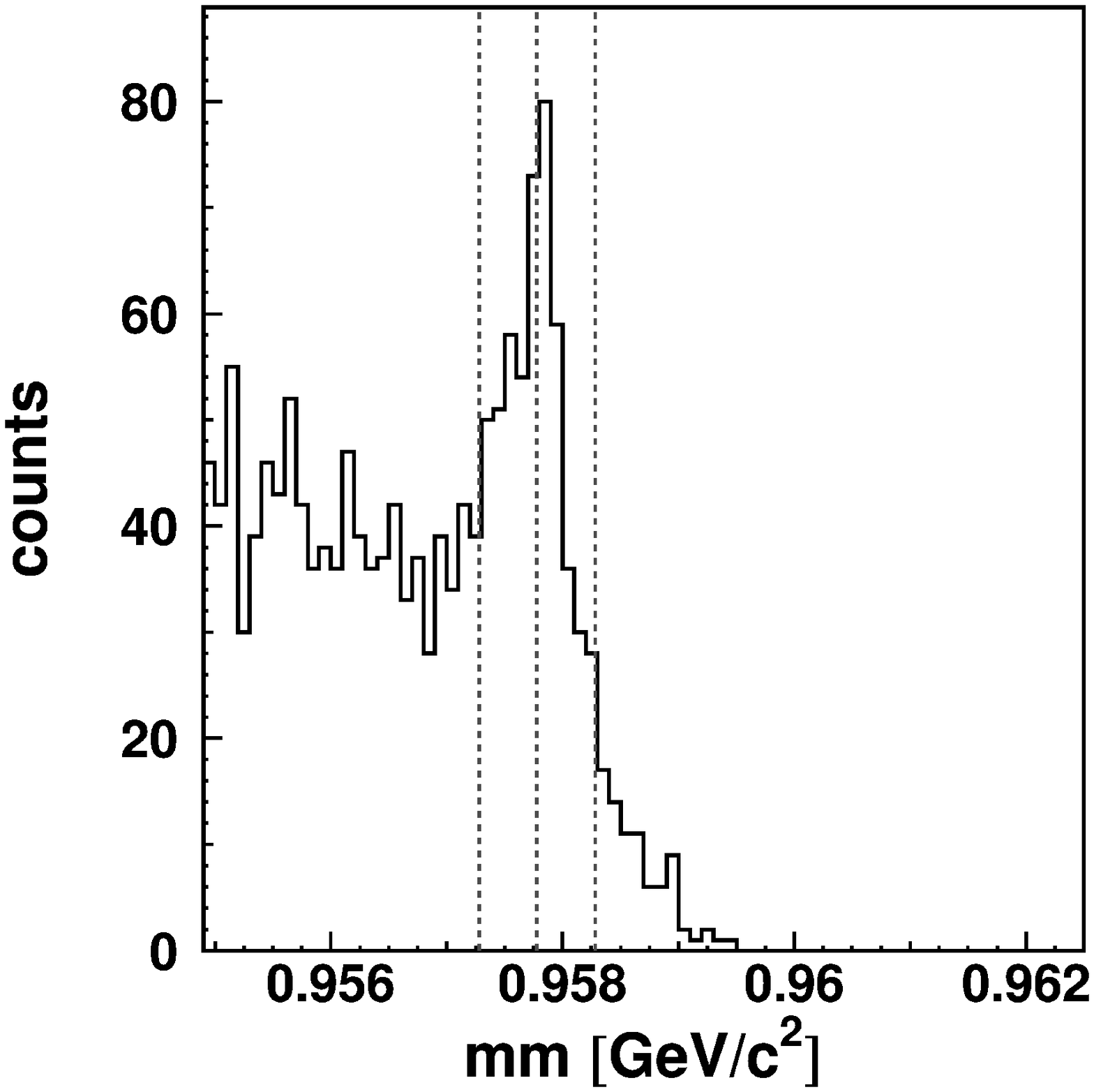}}
      }
      \makebox[0.33\textwidth][c]{
       \makebox[0.045\textwidth][c]{}
       \makebox[0.33\textwidth][c]{\includegraphics[width=0.25\textwidth]{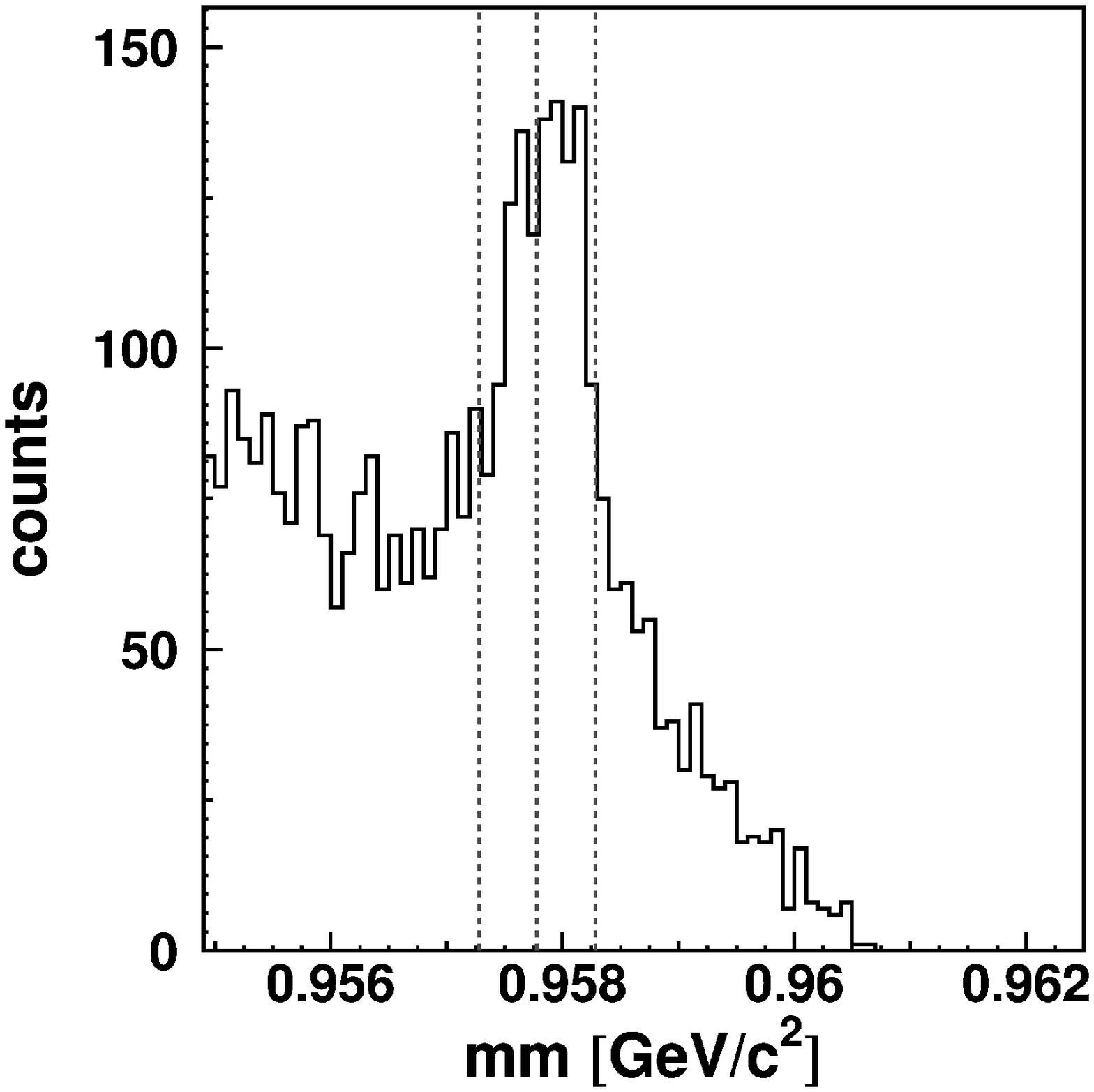}}
      }
      \makebox[0.33\textwidth][c]{
       \makebox[0.035\textwidth][c]{}
       \makebox[0.33\textwidth][c]{\includegraphics[width=0.25\textwidth]{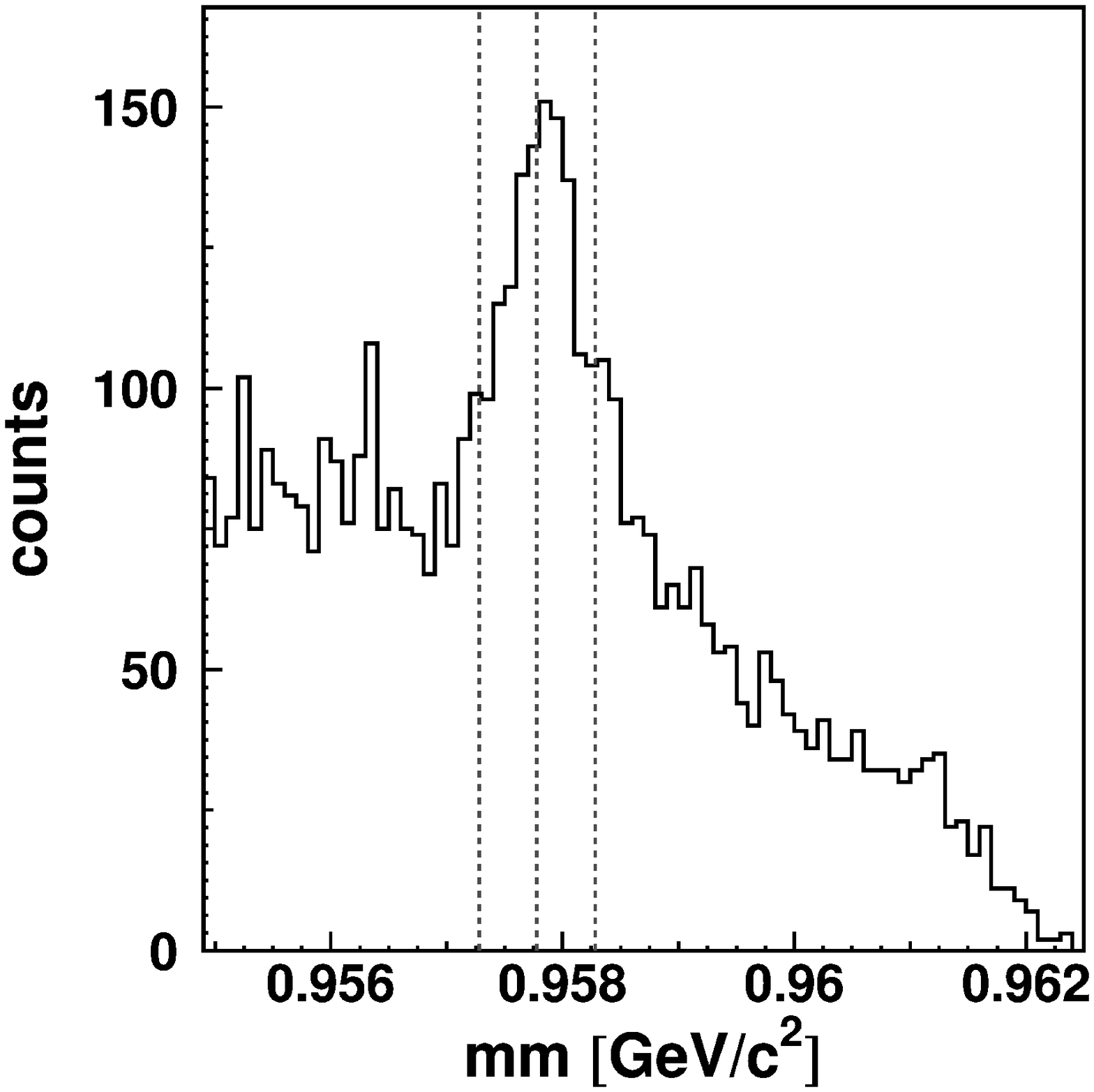}}
      }
     }
    }
    \makebox[0pt]{
     \raisebox{0.16\textheight}{
      \makebox[-0.95\textwidth][c]{}
      \makebox[0.0\textwidth][c]{{\bf a)}}
     }
    } 
    \makebox[0pt]{
     \raisebox{0.16\textheight}{
      \makebox[-0.27\textwidth][c]{}
      \makebox[0.0\textwidth][c]{{\bf b)}}
     }
    } 
    \makebox[0pt]{
     \raisebox{0.16\textheight}{
      \makebox[0.38\textwidth][c]{}
      \makebox[0.0\textwidth][c]{{\bf c)}}
     }
    } 
    \makebox[0pt]{
     \raisebox{-0.04\textheight}{
      \makebox[-0.99\textwidth][c]{}
      \makebox[0.0\textwidth][c]{{\bf d)}}
     }
    } 
    \makebox[0pt]{
     \raisebox{-0.04\textheight}{
      \makebox[-0.33\textwidth][c]{}
      \makebox[0.0\textwidth][c]{{\bf e)}}
     }
    } 
    \makebox[0pt]{
     \raisebox{-0.04\textheight}{
      \makebox[0.34\textwidth][c]{}
      \makebox[0.0\textwidth][c]{{\bf f)}}
     }
    } 
\caption{a)~Simulated FWHM of missing mass signal as a function of beam momentum above
           the threshold for the $\eta'$ meson creation in proton-proton collision.
           Squares:~target width 9~mm. Circles:~target width 1~mm~\cite{eryk}.
         $P_{beam}$ and $P_{th}$ denote the nominal beam momentum and the threshold momentum value
         for the $\eta'$ meson production in the $pp\to pp\eta'$ reaction.
         Plots from b) to e):~Missing mass spectra for the $pp\to ppX$ reaction measured at COSY--11
         detection setup at the beam momenta: a) 3211, b) 3213, c) 3214, d) 3218, e) 3224~MeV/c.
         Dashed lines indicate a  $\pm$0.5~MeV/c$^{2}$ band around the mass
         of the $\eta'$ meson.}
\label{szesc}
\end{figure}

An online analysis has  revealed a  signal
originating from the production of the $\eta^{\prime}$ meson
at each of the investigated beam momenta (Fig.~\ref{szesc}b-f).
As expected the width of the signal form the $\eta^{\prime}$ meson increases with increasing beam momentum.
The  width of the $\eta'$ signal determined closest to threshold
equals to approximatly 0.4~MeV (FWHM).
Taking into account that the width of the $\eta^{\prime}$ is around 0.2~MeV
we may estimate the achieved experimental resolution to be about 0.3~MeV 
just at the same order as the searched signal.  Presented spectra 
were obtained with preliminary calibration of the detection system and they will be
corrected for the possible broadening due to the changes 
of the beam optics which could caused the variations of
the beam momentum value at the order of 10$^{-5}$.
In order to enable such corrections
we have monitored various parameters which could influence the beam conditions
like current intensity in the COSY dipoles, the temperature of the cooling water of the magnets,
air temperature, humidity and barometric pressure inside COSY tunnel.
Independently, it will be possible to correct
the variation of the beam momentum based on
the distribution of the elasticaly scattered protons
measured simultaneously with the $pp\to pp\eta^{\prime}$ reaction.
Thus there is still a room for the improvement of the experimental resolution.
At present the off-line analysis of the data is in progress.

\begin{theacknowledgments}
We acknowledge the support of the
European Community-Research Infrastructure Activity
under the FP6 programme (Hadron Physics, N4:EtaMesonNet,
RII3-CT-2004-506078), the support
of the Polish Ministry of Science and Higher Education
under the grants
No. PB1060/P03/2004/26, 3240/H03/2006/31 and 1202/DFG/2007/03,
and  the support of the German Research Foundation (DFG).
\end{theacknowledgments}

\end{document}